\documentclass[12pt]{article}
\usepackage{tikz}
\usepackage{xcolor}
\usetikzlibrary{patterns, positioning, arrows}
\usepackage{graphicx} % Required for inserting images
\usepackage{amsmath,amsfonts,amssymb,mathtools}
\usepackage[nosort]{cite}
\usepackage{hyperref}
\usepackage{graphicx}
\usepackage{color}
\usepackage{epsfig}
\usepackage{psfrag}	
\usepackage{tikz}
\usepackage{cite}
\usepackage{slashed}
\usepackage{tensor}
\usepackage{listings}
\usepackage{ulem}
\usepackage[font=small,labelfont=bf,width=1\textwidth]{caption}
\usepackage{comment}
\usepackage[all]{xy}
\usepackage{multirow}
\usepackage{float}
\usepackage{cite}
\usepackage{color}

\usepackage{tocloft} % Layout of table of contents
\usepackage{ulem}
\usepackage{soul,xcolor}
\setstcolor{red}

\numberwithin{equation}{section}

\DeclareMathOperator{\llangle}{\big\langle\hspace{-1.2mm}\big\langle\hspace{-.5mm}}
\DeclareMathOperator{\rrangle}{\hspace{-.5mm}\big\rangle\hspace{-1.2mm}\big\rangle}

\DeclareMathOperator{\Li}{Li}

\newcommand{\bea}{\begin{eqnarray}}
\newcommand{\eea}{\end{eqnarray}}
\newcommand{\beq}{\begin{equation}}
\newcommand{\eeq}{\end{equation}}
\newcommand{\bal}{\begin{equation}\begin{aligned}}
\newcommand{\eal}{\end{aligned} \end{equation}}

\newcommand{\vev}[1]{{\left< {#1} \right>}}

\newcommand{\address}[1]{\vbox{\center\em#1}}
\renewcommand{\title}[1]{\vbox{\center\huge{#1}}\vspace{5mm}}

\newcommand{\cO}{{\mathcal O}}

\begin{document}

\setstcolor{red}

\begin{titlepage}
\begin{center}

\vspace*{10mm}

\title{Inversion and Integral Identities in dCFTs}

\vspace{7mm}

\renewcommand{\thefootnote}{$\alph{footnote}$}

%Nadav Drukker,%
%\textsuperscript{1,}
%\footnote{\href{mailto:nadav.drukker@gmail.com}
%{\tt nadav.drukker@gmail.com}}
%Andreas Stergiou,%
%\textsuperscript{1,}
%5\footnote{\href{mailto:andreas.stergiou@kcl.ac.uk}{\tt andreas.stergiou@kcl.ac.uk}}
%Ziwen Kong,%
%\textsuperscript{1,}
%\footnote{\href{mailto:ziwen.kong@kcl.ac.uk}
%{\tt ziwen.kong@kcl.ac.uk}}
Georgios Sakkas
\footnote{\href{mailto:georgios.sakkas@kcl.ac.uk}
{\tt georgios.sakkas@kcl.ac.uk}}

\vskip 2mm
\address{
Department of Mathematics, King's College London,
\\
The Strand, WC2R 2LS London, United-Kingdom}

\begin{abstract}
    This work derives an application from the identities of \cite{Kalmykov:2006pu} in order to invert four point functions in defect conformal field theories. For this, a recursion relation is established and the $O(N)$ model with a line defect is considered as a testing ground of this application. Specifically, the CFT data are calculated from inversion of tilt and displacement four point functions. The recursion relation enables efficient computation of hypergeometrics at order $\epsilon$ in the $\epsilon$-expansion, leading to the inversion of four point functions and the derivation of CFT data. The inversion method presented offers a faster alternative to traditional approaches using \cite{package1,package2}. The study also explores a general ansatz approach, assessing the algorithm's restrictiveness, and concludes by examining implications for the integral identity constraint of \cite{Drukker:2022dzs}, predicting corrections to OPE coefficients. 
\end{abstract}

\renewcommand{\thefootnote}{\arabic{footnote}}
\setcounter{footnote}{0}

\end{center}

\end{titlepage}

%\tableofcontents
%\newpage
%%%%%%%%%%%%%
\section{Introduction}
\label{sec:intro}
Conformal Field Theory stands as a cornerstone in the study of quantum field theories, providing a powerful framework to describe physical systems with scale invariance \cite{maldacena:1997re,Witten:1988hf,Aharony:1997an, David,Simmons-David, Dolan&Osborn_1,Dolan&Osborn_2,Dolan&Osborn_3,Osborn&Petkou }. Within the realm of CFT, understanding relationships between correlation functions, operator dimensions, and OPE coefficients is of paramount importance. This leads to the idea of inverting correlators to obtain CFT data. Given the four point function of some operator in a CFT, one can use it to extract from it CFT data, notable examples are \cite{Duffin:2017, Simon:2017,Liendo:2019,Liendo:2018ukf,Bianchi:2020hsz}. In this work, we present an application of the relations found in \cite{Kalmykov:2006pu}, and we use it to invert four point functions and obtain defect CFT data in defect conformal field theories. We take as an example the $O(N)$ model with a line defect in $d=4-\epsilon$ dimensions \cite{Andy , Gimenez-Grau:2022czc , Wilson:1972}. Due to operator mixing, we derive averaged CFT data (see Appendix \ref{Appendix_A}), from the four point functions of the tilt and displacement operators. This application involves the derivation of a recursion relation, that allows us to compute hypergeometrics exactly at order $\epsilon$ in $d=4-\epsilon$ dimensions. Then, with the use of the hypergeometrics and the explicit four point functions, we can derive the averaged CFT data by inversion. 

The study of defects in CFT is a very active area of research and has produced important results so far, some of them are \cite{Gliozzi:2015qsa, Drukker:2017dgn , Liendo:2012hy, Liendo:2018ukf, bianchi:2018zpb, Giombi:2018qox, Bianchi:2018scb, Giombi:2017cqn, Wang:2020xkc,Drukker:2020dcz,Herzog:2017kkj,Barrat:2021yvp,Barrat:2021tpn,Herzog:2017xha,Gromov:2022cgjp,Lemos:2017vnx}. In a CFT with a global symmetry $G$, when a line defect is introduced, the conformal group $SO(d+1,1)$ gets broken down to $SO(2,1)\times SO(d-1)$, it can happen that the line defect also breaks the global symmetry group down to $G'$. The breaking of the conformal group introduces the displacement operator, and the breaking of the global group introduces the tilt operator \cite{Gimenez-Grau:2022czc}. We will invert four point functions of these two operators obtaining the averaged defect CFT data. The defect deformations we will consider are of the form
\begin{align}
    S \xrightarrow{} S + h\int d\tau \phi_1(\tau) \,,
\end{align}
where $\tau$ is the coordinate on the line. This deformation will break the global symmetry group and introduce both the tilt and the displacement. The four point function of a generic operator $\mathrm{O}$ in the deformed theory will be given by
\begin{align}
\llangle\mathrm{O}(\tau_1)\mathrm{O}(\tau_2)\mathrm{O}(\tau_3)\mathrm{O}(\tau_4)\rrangle=\frac{\langle e^{h\int d\tau \phi_1(\tau)}\mathrm{O}(\tau_1)\mathrm{O}(\tau_2)\mathrm{O}(\tau_3)\mathrm{O}(\tau_4)\rangle}{\langle e^{h\int d\tau \phi_1(\tau)}\rangle} \,.
\end{align}

As discussed, the inversion of the four point functions we use here, is based on a recursion relation that obtains the $1d$ conformal block given by the hypergeometric function
$_2F_1(\Delta,\Delta;2\Delta;\xi)$, with $\xi$ the $1d$ conformal cross ratio, given by $\xi=\frac{\tau_{12}\tau_{34}}{\tau_{13}\tau_{24}}$, with $\tau_{12}=\tau_1-\tau_2$, for more on this see (Appendix \ref{conformal_block}). 
We will make use of important hypergeometric relations written in \cite{Kalmykov:2006pu} to derive the recursion relation, and then use it to obtain the hypergeometric in order to invert the four point function. We find that this algorithmic procedure is much faster than traditional methods of using Mathematica packages to expand the hypergeometrics as in \cite{package1,package2}.

We will also explore a general ansatz approach to the inversion, with the aim of understanding how restrictive this algorithmic method is. Finally, we explore the consequences of the above, to the integral identity constraint of \cite{Drukker:2022dzs}. The result is a constraint involving defect CFT data, that allows us to predict corrections of OPE coefficients. We begin our discussion with the recursion relation.
\section{Recursion}
\label{sec:recursion}
In this section we discuss the recursion relation, we will use this relation to derive the relevant hypergeometrics for the conformal blocks later on.\footnote{We would like to thank Andreas Stergiou for pointing out this paper \cite{Kalmykov:2006pu}} The equations we are going to need are \cite{Kalmykov:2006pu} 
\begin{align}
\label{identity_1}
    (c-a)(c-b) {\,}_2F_1(a,b;c+1;\xi) = ab(1-\xi) {\,}_2F_1(&a+1,b+1;c+1;\xi)\nonumber\\
    &-c(a+b-c) {\,}_2F_1(a,b;c;\xi) 
\end{align}
and 
\begin{align}
\label{der_equation}
    \frac{d}{d\xi}{\,}_2F_1(a,b;c;\xi)=\frac{ab}{c}{\,}_2F_1(a+1,b+1;c+1;\xi) \,.
\end{align}
The expression for $_2F_1(1+\gamma \varepsilon,1+\gamma \varepsilon;2+2\gamma \varepsilon;\xi)$ is given in \cite{Kalmykov:2006pu} and reads up to order $\varepsilon$ 
\begin{align}
_2F_1(1+\gamma \varepsilon,1+\gamma \varepsilon;2+2\gamma \varepsilon;\xi) = 
-\frac{\ln(1-\xi)}{\xi}-\frac{2\gamma\epsilon}{\xi}\Big(\ln(1-\xi)+\Li_2(\xi)\Big) + O(\varepsilon^2) \,.
\end{align}
This will be the hypergeometric we use to express all the others that we will need.
Let us start with the recursion that relates $_2F_1(n + \gamma\varepsilon,n + \gamma\varepsilon;n +1 + 2\gamma\varepsilon; \xi)$ with $_2F_1( 1 + \gamma\varepsilon , 1 + \gamma\varepsilon; 2 + 2\gamma\varepsilon; \xi)$ for $n$ a positive integer. With $(a)_n=\Gamma(a+n)/\Gamma(n)$ the Pochhammer symbol, this is given by 
\begin{align}
\label{first_step}
 _2F_1(n + \gamma\varepsilon,n + \gamma\varepsilon; n +1 + 2\gamma\varepsilon ,\xi)=\frac{(2+2\gamma\varepsilon)_{n-1}}{[(1+\gamma\varepsilon)_{n-1}]^2}\frac{d^{n-1}}{d\xi^{n-1}} {\,}_2F_1( 1 + \gamma\varepsilon , 1 + \gamma\varepsilon;2 + 2\gamma\varepsilon; \xi) \,.
\end{align}
One can deduce this from \eqref{der_equation}. Next, we can use \eqref{identity_1} with $a=b=n+\gamma\varepsilon$ and $c=n+m-1+2\gamma\varepsilon$, and we get
\begin{align}
    (m-1+\gamma\varepsilon)^2{\,}_2F_1(n+\gamma\varepsilon,n+\gamma\varepsilon;n+m+2\gamma\varepsilon;\xi)&=\nonumber\\
    &\hspace{-6.5cm}(n+\gamma\varepsilon)^2(1-\xi){\,}_2F_1(n+1+\gamma\varepsilon,n+1+\gamma\varepsilon;n+m+2\gamma\varepsilon;\xi)\nonumber\\
    &\hspace{-6.5cm}-(n+m-1+2\gamma\varepsilon)(n-m+1){\,}_2F_1(n+\gamma\varepsilon,n+\gamma\varepsilon;n+m-1+2\gamma\varepsilon;\xi) \,.
\end{align}
Using \eqref{der_equation} we find the recursion formula
\begin{align}
\label{recursion}
    _2F_1(n+\gamma\varepsilon,n+\gamma\varepsilon;n+m+2\gamma\varepsilon;\xi)&=\nonumber\\&\hspace{-5.5cm}
    \frac{(n+m-1+2\gamma\varepsilon)}{(m-1+\gamma\varepsilon)^2}\Big((1-\xi)\frac{d}{d\xi}-(n-m+1)\Big){\,}_2F_1(n+\gamma\varepsilon,n+\gamma\varepsilon;n+m-1+2\gamma\varepsilon;\xi) \,.
\end{align}
Now let's explain the algorithmic procedure. First given $n$ we find $_2F_1(n + \gamma\varepsilon,n + \gamma\varepsilon;n +1 + 2\gamma\varepsilon ;\xi)$ by using \eqref{first_step}, then we use \eqref{recursion} with initiating $m=2$ and up to $m=n$ and we print the resulting hypergeometric which is $_2F_1(n+\gamma\varepsilon,n+\gamma\varepsilon;2n+2\gamma\varepsilon;\xi)$. Since we have generated the relevant hypergeometrics with the use of the recursion relation, we can expand the four point function and match with the explicit result to find the CFT data, this is the inversion step. We explain in more detail in the next section.
\section{Analytic Defect Bootstrap}
\label{Analytic}
In this section we deploy the inversion algorithm, the goal is to find CFT data in defect theories. The example we will look at is the $O(N)$ model with a line defect. 

Let us start by writing the four point function of a general defect operator $O$ in terms of conformal blocks, the decomposition reads \cite{Dolan&Osborn_3} (and see appendix \ref{conformal_block})
\begin{align}
\label{4pt_decomposition}
  \llangle O_1(\tau_1)O_2(\tau_2)O_3(\tau_3)O_4(\tau_4)\rrangle=\frac{\mathcal{N}^2_O}{\tau_{12}^{2\Delta_O}\tau_{34}^{2\Delta_O}} \sum_{\Delta}\lambda_{OOO_\Delta}^2\xi^{\Delta}{\,}_2F_1(\Delta,\Delta;2\Delta;\xi) \,,
\end{align}
where $\Delta_O$ is the scaling dimension of the external operator and $\Delta=n+\gamma_n\epsilon$ is the dimension of the exchanged operator. The four point function will also be given by
\begin{align}
\label{4pt_general}
     \llangle O_1(\tau_1)O_2(\tau_2)O_3(\tau_3)O_4(\tau_4)\rrangle=\frac{\mathcal{N}^2_O}{\tau_{12}^{2\Delta_O}\tau_{34}^{2\Delta_O}}G(\xi)\,.
\end{align}
Where $G(\xi)$ is a known function. The inversion algorithm works as follows, first we use the recursion relation \eqref{recursion} to obtain hypergeometrics for up to the specified $n$. Second, we equate equations \eqref{4pt_decomposition} and \eqref{4pt_general} getting the equation
\begin{align}
    G(\xi)=\sum_{\Delta}\lambda_{OOO_\Delta}^2\xi^{\Delta}{\,}_2F_1(\Delta,\Delta;2\Delta;\xi) \,.
\end{align}
We expand both sides of this equation in powers of $\xi$ and match order by order terms fixing the CFT data. The advantage here is that we have produced the hypergeometrics exactly in $\xi$ with the use of the recursion, this works faster than packages that expand the hypergeometrics in series \cite{package1,package2} and allows one to go higher in $n$ as we will see. 

The CFT data $\lambda_{OOO_\Delta}^2$ and $\gamma_n$ are averaged CFT data with
\begin{align}
\lambda_{OOO_\Delta}=\lambda_{OOO_\Delta}^{(0)}+\epsilon\lambda_{OOO_\Delta}^{(1)} \,.
\end{align}
With the term averaged CFT data we mean that they are some sum of CFT data of operators that mix at order $\epsilon$ due to having the same classical scaling dimension (See Appendix \ref{Appendix_A} for details).\footnote{I would like to thank Andreas Stergiou for useful insights in this matter.} The calculations that follow produce results for these averages. We are able to disentangle operators and solve the mixing problem for the case of the lowest lying singlet operators in the $O(N)$ model with a line defect, with $\Delta_{S1}=\Delta_{S2}=2$.
\subsection{$O(N)$ Model with a line}
Let us start this section by reviewing the $O(N)$ model with a line defect in $d=4-\epsilon$. The action for this theory is the following
\begin{align}
    S=\int d^dx\Big( \Big(\frac{1}{2}\partial_{\mu}\phi_A\Big)^2+\frac{\lambda_0}{4!}\Big(\phi_A^2)\Big)^2\Big)+h_0\int_{-\infty}^{\infty} d\tau\phi_1(\tau) ,\quad A=1,\dots,N,
\end{align}
where $\tau$ is the coordinate on the line. The values of the couplings at the fixed point are \cite{Gimenez-Grau:2022czc}, \cite{Cuomo:2022zm}
\begin{align}
   \label{Couplings_fixed_point}
   \frac{\lambda_*}{(4\pi)^2}=\frac{3\epsilon}{N+8}+O(\epsilon^2) \quad h^2_*=N+8+\epsilon\frac{4N^2+45N+170}{2(N+8)}+O(\epsilon^2) \,.
\end{align}
The defect breaks $O(N)\xrightarrow{}O(N-1)$ , this breaking introduces the tilt operator $t_i$ that appears in the broken conservation current equation 
\begin{align}
\label{tilt_current_eq}
    \partial_{\mu}J^{\mu}_i(x_{\perp},\tau)=t_i(\tau)\delta^{(d-1)}(x_{\perp})\,,
\end{align}
where $x_{\perp}$ are the transverse directions to the defect and $i=1,\dots,N-1$. The introduction of the defect breaks also the conformal group $SO(d+1,1)\xrightarrow{}SO(2,1)\times SO(d-1)$, this breaking introduces the displacement operator, more specifically the breaking of translations does. The displacement operator appears in the stress tensor conservation equation
\begin{align}
    \partial_{\mu}T^{\mu a}(x_{\perp},\tau)=D^{a}(\tau)\delta^{(d-1)}(x_{\perp})\,,
\end{align}
with $a=1,\dots,d-1$. Both operators have protected scaling dimension, the tilt $\Delta_t=1$, and the displacement $\Delta_D=2$. 

We should give now some further details about the spectrum of the theory as it will be useful for the analysis later on. The tilt is the lowest lying operator with scaling dimension exactly $\Delta_t=1$, the scalar $\phi_1$ that couples to the line is the second lowest lying operator, with scaling dimension given in the $\epsilon$-expansion by \cite{Gimenez-Grau:2022czc}
\begin{align}
\label{phi_1_sclaing}
    \Delta_{\phi_1}=1+\epsilon-\frac{3N^2+49N+194}{2(N+8)^2}\epsilon^2 +O(\epsilon^3)\,.
\end{align}
We now look at the OPE of tilts in the singlet, traceless-symmetric and anti-symmetric channels, the lowest lying operators are
\begin{align}
    \label{singlet_OPE_tilts}
    (t_i\times t_j)_S =\phi_1+S1+S2+\dots \,,
\end{align}
the operators $S1,S2$ can be realized in perturbation theory as a linear combination of $\phi_1^2$ and $\phi_A^2$ and their scaling dimensions have been calculated by the use of Feynman diagram techniques in \cite{Gimenez-Grau:2022czc}
\begin{align}
\label{S1,S2_scalings}
    \Delta_{S1,S2}=2+\epsilon\frac{3N+20\pm\sqrt{N^2+40N+320}}{2(N+8)}+O(\epsilon^2)\,.
\end{align}
The traceless-symmetric channel reads
\begin{align}
\label{Traceless-symmetric_OPE_tilts}
    (t_i\times t_j)_T=T_{ij}+\dots \,,
\end{align}
with $T_{ij}$ being the lowest lying operator in this channel with scaling dimension \cite{Gimenez-Grau:2022czc}
\begin{align}
    \Delta_T=2+\frac{2\epsilon}{N+8}+O(\epsilon^2)\,.
\end{align}
Lastly, the anti-symmetric channel 
\begin{align}
    (t_i\times t_j)_A=A_{ij}+\dots \,,
\end{align}
with $A_{ij}$ being the lowest lying operator in this channel with scaling dimension \cite{Gimenez-Grau:2022czc}
\begin{align}
    \Delta_A=3+O(\epsilon^2) \,.
\end{align}
The two point functions of tilts and displacements were also calculated in \cite{Gimenez-Grau:2022czc}, and they are given by
\begin{align}
    \llangle t_i(\tau_1)t_j(\tau_2)\rrangle=\frac{\mathcal{N}_t^2\delta_{ij}}{\tau_{12}^2\tau_{34}^2},\quad \llangle D_a(\tau_1)D_b(\tau_2)\rrangle=\frac{\mathcal{N}_D^2\delta_{ab}}{\tau_{12}^4\tau_{34}^4}\,,
\end{align}
with $\tau_{12}=\tau_1-\tau_2$, and the normalizations given by \cite{Gimenez-Grau:2022czc}
\begin{align}
\label{Normalizations}
    \mathcal{N}_t^2=h^2\kappa\Big(1-\frac{(1+\gamma_E+\log\pi)\epsilon}{2}\Big)+O(\epsilon^2),\quad \mathcal{N}_D^2=2\kappa\Big(1-\frac{4+3(1+\gamma_E+\log\pi)\epsilon}{6}\Big)+O(\epsilon^2)\,.
\end{align}
$\kappa$ here is given by $\kappa=\frac{\Gamma(\frac{d}{2})}{2\pi^{\frac{d}{2}}(d-2)}$.

In what follows we compute averaged CFT data for the higher dimensional operators that appear in all three of the above OPE's. We also compute data for operators appearing in the displacement channels (channels $S,T,A)$. Note, since operators with higher scaling dimensions mix, we compute here only averaged data of them. In the case of $S1,S2$ we can explicitly disentangle them and retrieve the result \eqref{S1,S2_scalings} of \cite{Gimenez-Grau:2022czc}, we perform this computation in subsection \ref{Mixed_scalar/tilt}.
\subsubsection{Tilts}
In this subsection we discuss the inversion of the four point function of tilts at order $\epsilon$, the four point function at order $\epsilon$ was calculated in \cite{Gimenez-Grau:2022czc} with the use of Feynman diagram techniques
\begin{align}
\label{4pt_tilt}
    \llangle t_i(\tau_1)t_j(\tau_2)t_k(\tau_3)t_l(\tau_4)\rrangle=\frac{\mathcal{N}_t^2}{\tau_{12}^2\tau_{34}^2}\Big(&\delta_{ij}\delta_{kl}+\delta_{ik}\delta_{jl}\xi^2+\delta_{il}\delta_{jk}\Big(\frac{\xi}{1-\xi}\Big)^2\nonumber\\&+\frac{2\epsilon I(\xi)}{N+8}(\delta_{ij}\delta_{kl}+\delta_{il}\delta_{jk}+\delta_{ik}\delta_{jl})\Big) \,.
\end{align}
With $I(\xi)$ given in \cite{Gimenez-Grau:2022czc}
\begin{align}
    I(\xi)=\xi\log(1-\xi)+\frac{\xi^2}{1-\xi}\log\xi \,,
\end{align}
and $\xi$ the 1d conformal cross-ratio
\begin{align}
    \xi=\frac{\tau_{12}\tau_{34}}{\tau_{13}\tau_{24}}\,.
\end{align}
An alternative representation of the four point function in terms of the singlet, traceless-symmetric and anti-symmetric channels is the following
\begin{align}
\label{4ptSAT}
    \llangle t_i(\tau_1)t_j(\tau_2)t_k(\tau_3)t_l(\tau_4)\rrangle=\frac{\mathcal{N}_t^2}{\tau_{12}^2\tau_{34}^2}&\Big(\delta_{ij}\delta_{kl}G_t^S(\xi)+(\delta_{il}\delta_{jk}-\delta_{ik}\delta_{jl})G_t^A(\xi)\nonumber\\&+\Big(\frac{1}{2}\delta_{ik}\delta_{jl}+\frac{1}{2}\delta_{il}\delta_{jk}-\frac{\delta_{ij}\delta_{kl}}{N-1}\Big)G_t^T(\xi)\Big) \,.
\end{align}
Using \eqref{4pt_tilt} the $G_t^S,G_t^A,G_t^T$ functions are given by
\begin{align}
    &G_t^S(\xi)=1+\frac{1}{N-1}\Big(\xi^2+\Big(\frac{\xi}{1-\xi}\Big)^2\Big)+\frac{2(N+1)\epsilon}{(N-1)(N+8)}I(\xi)\,,\\
    &G_t^A(\xi)=\frac{1}{2}\Big(\Big(\frac{\xi}{1-\xi}\Big)^2-\xi^2\Big)\,,\\
    &G_t^T(\xi)=\Big(\xi^2+\Big(\frac{\xi}{1-\xi}\Big)^2\Big)+ \frac{4\epsilon}{N+8} I(\xi) \,.
\end{align}
We can do the inversion of the above functions and obtain the averaged CFT data for every channel. First, we will consider the singlet channel. We use the recursion relation to collect all the relevant hypergeometrics up to the specified $n$, and then find the averaged CFT data. For the singlet channel up to $n=4$ the results are
\begin{align}
    & \lambda_{ttS}^{(0)}[1]=0,\quad \gamma_{S}[2]=\frac{N+1}{N+8},\quad\lambda_{ttS}^{(0)}[2]=-\frac{\sqrt{2}}{\sqrt{N-1}},\quad\lambda_{ttS}^{(1)}[2]=\frac{N+1}{\sqrt{2(N-1)}(N+8)},\nonumber\\&\quad \lambda_{ttS}^{(0)}[3]=0,\quad \lambda_{ttS}^{(0)}[4]=-\sqrt{\frac{6}{5(N-1)}},\quad \lambda_{ttS}^{(1)}[4]=\frac{37(N+1)}{6\sqrt{30(N-1)}(N+8)},\quad \gamma_{S}[4]=\frac{N+1}{6(N+8)}\,.
\end{align}
In the Mathematica file we have calculated averaged CFT data for the singlet channel up to $n=10$. Note that the algorithm finds several solutions, we choose to display the first one. The reason for this, is that OPE coefficients appear squared and therefore they have two solutions, different choices for the sign of the OPE coefficients results into different inversion solutions. 

We proceed with the traceless symmetric channel, we get results for $n=4$ 
\begin{align}
 &\lambda_{ttT}^{(0)}[1]=0,\quad \gamma_{T}[2]= \frac{2}{N+8},\quad  \lambda_{ttT}^{(0)}[2]=-\sqrt{2},\quad \lambda_{ttT}^{(1)}[2]=\frac{\sqrt{2}}{N+8},\quad \lambda_{tt T}^{(0)}[3]=0, \\& \lambda_{ttT}^{(0)}[4]=-\sqrt{\frac{6}{5}},\quad \lambda_{ttT}^{(1)}[4]=\frac{37}{30\sqrt{30}(N+8)},\quad \gamma_{T}[4]=\frac{1}{3(N+8)}\,.
\end{align}
In the Mathematica file we've managed to get results for the traceles symmetric channel up to $n=10$.

The remaining channel to invert is the anti-symmetric one, the results for $n=5$ read
\begin{align}
  &\lambda_{ttA}^{(0)}[1]=\lambda_{ttA}^{(0)}[2]=0,\quad\lambda_{ttA}^{(0)}[3]=-1 ,\quad \lambda_{ttA}^{(1)}=0 ,\quad \gamma_{A}[3]=0 ,\\&\lambda_{ttA}^{(0)}[4]=0,\quad \lambda_{ttA}^{(0)}[5]=-\sqrt{\frac{2}{7}},\quad \lambda_{ttA}^{(1)}[5]=0,\quad \gamma_{A}[5]=0 \,.
\end{align}
In the Mathematica file we've managed to get results up to $n=11$.
\subsubsection{Displacements}
We report here results for inverting the displacement four point function in the $O(N)$ model with a line defect. First, let us start by giving the explicit four point function as it was derived in \cite{Gimenez-Grau:2022czc}
\begin{align}
    \llangle D_{a}(\tau_1)D_{b}(\tau_2)D_{c}(\tau_3)D_{d}(\tau_4)\rrangle=&\frac{\mathcal{N}_{D}^2}{\tau_{12}^4\tau_{34}^4}\Big(\delta_{ab}\delta_{cd}+\delta_{ac}\delta_{bd}\xi^4+\delta_{ad}\delta_{bc}\Big(\frac{\xi}{1-\xi}\Big)^4\\&+\frac{2\epsilon I_1(\xi)}{5(N+8)}(\delta_{ab}\delta_{cd}+\delta_{ac}\delta_{bd}+\delta_{ad}\delta_{bc})\Big)\,,
\end{align}
with $I_1(\xi)$ given by
\begin{align}
    I_1(\xi)=\frac{(\xi^2-\xi+1)\xi^2}{(1-\xi)^2}+\frac{1}{2}(2\xi^2+\xi+2)\xi\log(1-\xi)+\frac{(2\xi^2-5\xi+5)\xi^4}{2(1-\xi)^3}\log\xi\,.
\end{align}
We can re-write this using a decomposition in terms of the singlet, traceless-symmetric and anti-symmetric channel
\begin{align}
    \llangle D_{a}(\tau_1)D_{b}(\tau_2)D_{c}(\tau_3)D_{d}(\tau_4)\rrangle = &\frac{\mathcal{N}_{D}^2}{\tau_{12}^4\tau_{34}^4}\Big(\delta_{ab}\delta_{cd}G_D^S(\xi)+(\delta_{ad}\delta_{bc}-\delta_{ac}\delta_{bd})G_D^A(\xi)\\&+\Big(\frac{1}{2}\delta_{ac}\delta_{bd}+\frac{1}{2}\delta_{ad}\delta_{bc}-\frac{\delta_{ab}\delta_{cd}}{d-1}\Big)G_D^T(\xi)\Big) \,,
\end{align}
using this we can find $G_D^S,G_D^A,G_D^T$
\begin{align}
&G_D^S(\xi)=1+\frac{\Big(\Big(\frac{\xi}{1-\xi}\Big)^4+\xi^4\Big)}{d-1}+\frac{2(d+1)\epsilon I_1(\xi)}{5(N+8)(d-1)} \,,\\
    &G_D^T(\xi)=\xi^4+\Big(\frac{\xi}{1-\xi}\Big)^4+\frac{4\epsilon I_1(\xi)}{5(N+8)}\,,\\
    &G_D^A(\xi)=\frac{1}{2}\Big(-\xi^4+\Big(\frac{\xi}{1-\xi}\Big)^4\Big)\,.
\end{align}
Expanding in $d=4-\epsilon$ we get
\begin{align}
\label{Displacement_singlet}
     &G_D^S(\xi)=1+\frac{1}{3}\Big(\Big(\frac{\xi}{1-\xi}\Big)^4+\xi^4\Big)+\frac{\epsilon}{9}\Big(\Big(\frac{\xi}{1-\xi}\Big)^4+\xi^4\Big)+\frac{2\epsilon I_1(\xi)}{3(N+8)} \,,\\ \label{Displacement_traceles}
    &G_D^T(\xi)=\xi^4+\Big(\frac{\xi}{1-\xi}\Big)^4+\frac{4\epsilon I_1(\xi)}{5(N+8)}\,,\\
    \label{displacement_anti_symmetric}
    &G_D^A(\xi)=\frac{1}{2}\Big(-\xi^4+\Big(\frac{\xi}{1-\xi}\Big)^4\Big)\,.
\end{align}
Using as input equations\eqref{Displacement_singlet},\eqref{Displacement_traceles},\eqref{displacement_anti_symmetric} in the inversion algorithm we can extract averaged CFT data for the displacement. We report here the calculation for $n=4$, first for the singlet channel
\begin{align}
    &\lambda_{DDS}^{(0)}[1]= \lambda_{DDS}^{(0)}[2]= \lambda_{DDS}^{(0)}[3]=0,\quad\lambda_{DDS}^{(0)}[4]=-\sqrt{\frac{2}{3}},\quad\lambda_{DDS}^{(1)}[4]=-\frac{37+4N}{12\sqrt{6}(N+8)},\\& \gamma_{DS}[4]=\frac{5}{2(N+8)}\,.
\end{align}
In the Mathematica file we have managed to get averaged CFT data up to $n=10$.
For the traceles symmetric channel we find up to $n=4$
\begin{align}
    &\lambda_{DDT}^{(0)}[1]= \lambda_{DDT}^{(0)}[2]= \lambda_{DDT}^{(0)}[3]=0,\quad\lambda_{DDT}^{(0)}[4]=-\sqrt{2},\quad\lambda_{DDT}^{(1)}[4]=-\frac{1}{6\sqrt{2}(N+8)},\\& \gamma_{DT}[4]=\frac{1}{(N+8)}\,.
\end{align}
Simillarly, we can push this up to $n=10$ in the Mathematica file.

For the anti-symmetric channel we get for $n=5$
\begin{align}
    \lambda_{DDA}^{(0)}[1]=\lambda_{DDA}^{(0)}[2]=\lambda_{DDA}^{(0)}[3]=\lambda_{DDA}^{(0)}[4]=\lambda_{DDA}^{(1)}[5]=0,\quad \lambda_{DDA}^{(0)}[5]=-\sqrt{2},\quad \gamma_{DA}[5]=0\,,
\end{align}
and once again we can get results up to $n=11$ for this channel. 
\subsubsection{Mixed Scalar/Tilt}
\label{Mixed_scalar/tilt}
We wish now to disentangle the lowest lying singlet operators $(S1,S2)$ with classical scaling dimensions $\Delta_{S1}=\Delta_{S2}=2$ in the $O(N)$ model with a line defect\footnote{I would like to thank Prof. Nadav Drukker for pointing out this direction and suggesting the problem.}. To do this we will need the four point function of the scalars $\phi_1$ and the mixed four point function of both $\phi_1$ and $t_i$. Both were derived in \cite{Gimenez-Grau:2022czc} and read respectively
\begin{align}
\label{4pt_phi}
    \llangle \phi_1(\tau_1)\phi_1(\tau_2)\phi_1(\tau_3)\phi_1(\tau_4)\rrangle = \frac{\mathcal{N}_{\phi_1}^2}{\tau_{12}^{2\Delta_{\phi_1}}\tau_{34}^{2\Delta_{\phi_1}}}\Big(1+\xi^{2\Delta_{\phi_1}}+\Big(\frac{\xi}{1-\xi}\Big)^{2\Delta_{\phi_1}}+\frac{6\epsilon I(\xi)}{N+8}+O(\epsilon^2)\Big)
\end{align}
and
\begin{align}
\label{4pt_Mixed}
    \llangle \phi_1(\tau_1)\phi_1(\tau_2)t_i(\tau_3)t_j(\tau_4)\rrangle = \frac{\mathcal{N}_{\phi_1}\mathcal{N}_{t}}{\tau_{12}^{2\Delta_{\phi_1}}\tau_{34}^{2}}\Big(\delta_{ij}+\frac{2\epsilon\delta_{ij} I(\xi)}{N+8}+O(\epsilon^2)\Big) \,.
\end{align}
We can now do the inversion, first we consider the four point function of $\phi_1$'s and find the following constraints
\begin{align}
\label{scalars_degeneracy_1}
    &\lambda_{\phi_1\phi_1 S1}^2+\lambda_{\phi_1\phi_1 S2}^2=2-\frac{6\epsilon}{N+8} \,,\\\label{scalars_degeneracy_2}
    &(\Delta_{S1}-2)\lambda_{\phi_1\phi_1 S1}^2+(\Delta_{S2}-2)\lambda_{\phi_1\phi_1 S2}^2=\epsilon\frac{38+4N}{N+8}\,,
\end{align}
with 
\begin{align}
    \Delta_{S1}=2+\epsilon\gamma_{S1} ,\quad \Delta_{S2}=2+\epsilon\gamma_{S2} 
\end{align}
and
\begin{align}
    &\lambda_{\phi_1\phi_1 S1}=\lambda_{\phi_1\phi_1 S1}^{(0)}+\epsilon\lambda_{\phi_1\phi_1 S1}^{(1)} \,,\\
    &\lambda_{\phi_1\phi_1 S2}=\lambda_{\phi_1\phi_1 S2}^{(0)}+\epsilon\lambda_{\phi_1\phi_1 S2}^{(1)} \,.
\end{align}
Similarly, we can do the same for the mixed four point function and we find
\begin{align}
\label{mixed_degeneracy_1}
&\lambda_{\phi_1\phi_1 S1}\lambda_{tt S1}+\lambda_{\phi_1\phi_1 S2}\lambda_{tt S2}=-\epsilon\frac{2}{N+8} \,,\\ \label{mixed_degeneracy_2}
&(\Delta_{S1}-2)\lambda_{\phi_1\phi_1 S1}\lambda_{tt S1}+(\Delta_{S2}-2)\lambda_{\phi_1\phi_1 S2}\lambda_{tt S2}=\epsilon\frac{2}{N+8} \,.
\end{align}
From the singlet tilt four point function we get 
\begin{align}
\label{tilt_degeneracy_1}
    &\lambda_{ttS1}^2+\lambda_{ttS2}^2=\frac{2}{N-1}-\epsilon\frac{2(N+1)}{(N-1)(N+8)}\,,\\ \label{tilt_degeneracy_2}
    &(\Delta_{S1}-2)\lambda_{ttS1}^2+(\Delta_{S2}-2)\lambda_{ttS2}^2=\epsilon\frac{2(N+1)}{(N-1)(N+8)} \,.
\end{align}
Solving now equations \eqref{scalars_degeneracy_1}, \eqref{scalars_degeneracy_2},\eqref{mixed_degeneracy_1}, \eqref{mixed_degeneracy_2}, \eqref{tilt_degeneracy_1},\eqref{tilt_degeneracy_2} gives the anomalous dimensions 
\begin{align}
\label{anomalous_dimensions_S1,2}
    \gamma_{S1}=\frac{3N+20+\sqrt{320+40N+N^2}}{2(N+8)},\quad \gamma_{S2}=\frac{3N+20-\sqrt{320+40N+N^2}}{2(N+8)} \,,
\end{align}
and the OPE coefficients at order $\epsilon^0$
\begin{align}
\label{OPE_coefficients_S1,2}
    &(\lambda_{\phi_1\phi_1 S1,S2}^{(0)})^2=1\pm\frac{18+N}{\sqrt{320+40N+N^2}}\,,\\&  (\lambda_{ttS1}^{(0)})^2=\frac{4}{320+40N+N^2+(18+N)\sqrt{320+40N+N^2}}\,,\\
    &  (\lambda_{ttS2}^{(0)})^2=\frac{1}{N-1}\Big(1+\frac{18+N}{\sqrt{320+40N+N^2}}\Big)\,.
\end{align}
This solves the mixing problem at this order in $\epsilon$ and disentangles the operators $S1,S2$. This was also found in \cite{Gimenez-Grau:2022czc} by calculating Feynman diagrams, here we rederived it with the use of the inversion algorithm alone.
\subsubsection{General Ansatz}
\label{general_ansatz}
The derivations of the CFT data so far were performed by taking as input the four point function and applying the inversion algorithm. In this section, we assume a general ansatz for the four point function of the tilts in the $O(N)$ defect model and investigate the consequences of the inversion algorithm along with the crossing equations to these ansatzes.
The general ansatzes for the singlet, traceless-symmetric and anti-symmetric channels that we will assume are the following
\begin{align}
\label{Ansatz_1}
    &G_{At}^S(\xi)=1+\xi^2+\frac{1}{N-1}\Big(\frac{\xi}{1-\xi}\Big)^2+\frac{\xi}{1-\xi}\Big(a_S \xi\log\xi+b_S (1-\xi)\log(1-\xi)\Big)\,,\\ \label{Ansatz_2}
    &G_{At}^T(\xi)=\xi^2+\Big(\frac{\xi}{1-\xi}\Big)^2+\frac{\xi}{1-\xi}\Big(a_T \xi\log\xi+b_T (1-\xi)\log(1-\xi)\Big)\,,\\ \label{Ansatz_3}
    &G_{At}^A(\xi)=\frac{1}{2}\Big(-\xi^2+\Big(\frac{\xi}{1-\xi}\Big)^2\Big)+\frac{\xi}{1-\xi}\Big(a_A \xi\log\xi+b_A (1-\xi)\log(1-\xi)\Big)\,.
\end{align}
The crossing equations are
\begin{align}
\label{crossing_1}
    \frac{G_{At}^T(\xi)}{2}-G_{At}^A(\xi)=\Big(\frac{\xi}{1-\xi}\Big)^2\Big(\frac{G_{At}^T(1-\xi)}{2}-G_{At}^A(1-\xi)\Big) \,,
\end{align}
and
\begin{align}
\label{crossing_2}
    G_{At}^S(\xi)-\frac{1}{N-1}G_{At}^T(\xi)=\Big(\frac{\xi}{1-\xi}\Big)^2\Big(\frac{G_{At}^T(1-\xi)}{2}+G_{At}^A(1-\xi)\Big)\,.
\end{align}
Using the crossing equations and the ansatzes \eqref{Ansatz_1},\eqref{Ansatz_2},\eqref{Ansatz_3} we find 3 constraints on the 6 unknown parameters
\begin{align}
\label{crossing_constraints_1}
    &b_T =2(b_A-a_A)+a_T,\quad  a_S=2b_A-a_A+a_T\frac{N+1}{2(N-1)},\\ \label{crossing_constraints_2}&b_S=\frac{2b_A}{N-1}+\frac{a_A(N-3)}{N-1}+a_T\frac{N+1}{2(N-1)} \,.
\end{align}
Applying the inversion algorithm to the anti-symmetric channel \eqref{Ansatz_3} we find that 
\begin{align}
    a_A=b_A=0 \,,
\end{align}
using this along with the 3 constraints \eqref{crossing_constraints_1},\eqref{crossing_constraints_2} from crossing, we find
\begin{align}
    b_T=a_T,\quad a_S=b_S=a_T\frac{N+1}{2(N-1)}\,.
\end{align}
Hence, we are left with only one undetermined parameter $a_T$. Applying the inversion algorithm now to the singlet and traceless-symmetric channels we find respectively
\begin{align}
   a_S=b_S=\frac{2\gamma_S[2]}{N-1},\quad \lambda_{ttS}^{(0)}[1]=0,\quad\lambda_{ttS}^{(0)}[2]=-\sqrt{\frac{2}{N-1}},\quad \lambda_{ttS}^{(1)}[2]=\frac{\gamma_S[2]}{\sqrt{2(N-1)}}\,.
\end{align}
and
\begin{align}
    a_T=b_T=2\gamma_T[2],\quad \lambda_{ttT}^{(0)}[1]=0,\quad \lambda_{ttT}^{(0)}[2]=-\sqrt{2},\quad \lambda_{ttT}^{(1)}[2]=\frac{\gamma_T[2]}{\sqrt{2}}\,.
\end{align}
We notice that we can't find explicitly the unknown parameter $a_T$, it is only at the cost of knowing one more anomalous dimension either $\gamma_S[2]$ or $\gamma_T[2]$, then and only then we can completely determine the 6 unknown parameters and recover the explicit results for the four point function of the tilts. So, we collect here the results of the above analysis 
\begin{align}
\label{result_1_g}
    &G_{At}^S(\xi)=1+\xi^2+\frac{1}{N-1}\Big(\frac{\xi}{1-\xi}\Big)^2+a_T\frac{\xi}{1-\xi}\frac{N+1}{2(N-1)}\Big( \xi\log\xi+(1-\xi)\log(1-\xi)\Big)\,,\\  \label{result_2_g}
    &G_{At}^T(\xi)=\xi^2+\Big(\frac{\xi}{1-\xi}\Big)^2+a_T\frac{\xi}{1-\xi}\Big( \xi\log\xi+ (1-\xi)\log(1-\xi)\Big)\,,\\  \label{result_3_g}
    &G_{At}^A(\xi)=\frac{1}{2}\Big(-\xi^2+\Big(\frac{\xi}{1-\xi}\Big)^2\Big)\,.
\end{align}
In the next section we consider the integral identities of \cite{Drukker:2022dzs} as a potential solution to identifying the parameter $a_T$. We find that due to a cancellation occurring in the integral this is not possible, but it allows instead the prediction of the order $\epsilon^2$ correction to the $\lambda_{tt\phi_1}$ OPE coefficient. 
\section{Integral Identities}
In this section we examine the consequences of the general ansatz analysis to the integral identities of \cite{Drukker:2022dzs}. We find a constraint involving defect CFT data of different orders in the $\epsilon$-expansion. This allows us to predict the order $\epsilon^2$ correction to the OPE coefficient $\lambda_{tt\phi_1}$ in the $O(N)$ model with a line defect. Moreover, we derive simillar constraints for defect CFT data in the chiral Heisenberg model and the $O(n)\times O(m)$ model using the fact that the four point functions of tilts in these models have the same analytic form as the one in the $O(N)$ model with a line defect.
\subsection{Review of Integral Identities}
In a conformal field theory, exactly marginal operators allow for continuous deformations of the theory, forming a space of CFTs known as the 
conformal manifold. In a $d$ dimensional CFT, marginal operators $\cO_i$ have scaling dimension 
$d$. If the CFT has an action, the deformations can be written as
\beq
\label{deformation}
S\to S+\int\lambda^i\cO_i\, d^dx\,,
\eeq
where the parameters $\lambda^i$ are local coordinates on the conformal 
manifold. Correlation functions of any operators $\varphi_a$ in the deformed 
theory are then written simply with the extra insertion of 
the exponential of the integral in \eqref{deformation}
\beq
\label{deformed}
\vev{\varphi_{a_1}\cdots\varphi_{a_n}}_{\lambda^i}
=\vev{e^{-\int\lambda^i\cO_i\, d^dx}\varphi_{a_1}\cdots\varphi_{a_n}}_{0}\,,
\eeq
with subscript 0 indicates the undeformed theory.

The breaking of a global symmetry gives rise to tilts $t_i$, these operators are marginal defect operators and as we discussed in section \ref{Analytic} they satisfy
\begin{align}
\label{broken_current_intro}
    \partial_{\mu}J^{\mu i}(x_{\perp},\tau)=t^i(\tau) \delta^{d-p}(x_{\perp}) \,,
\end{align}
where $p$ is the dimesnion of the defect. Their two-point function is given by
\beq
\llangle t_i(\tau) t_{j}(0)\rrangle =
\frac{\mathcal{C}_t^2 \delta_{ij}}{ |\tau|^{2p}}\,.
\eeq
This leads naturally to consider a defect conformal manifold, and 
with the usual rescaling of the operator at infinity, it has 
the Zamolodchikov metric~\cite{Zamolodchikov:1986gt} 
\beq
\label{Nt}
g_{ij}=\llangle t_{i}(\infty) t_{j}(0)\rrangle =
\mathcal{C}_t^2 \delta_{ij}\,,
\qquad
t_i(\infty)\equiv\lim_{\tau\to\infty}|\tau|^{2p}t_i(x_\parallel)\,.
\eeq

While the metric is locally flat, 
if the theory has global group $G$, broken by the defect to $G'$, the full 
defect conformal manifold is $G/G'$ \cite{Drukker:2022dzs}. Furthermore, 
the size of this manifold is set by $\mathcal{C}_t$.

This point of view 
allows one to find non-trivial identities on integrated correlators. The defect analog 
of \eqref{deformed} is
\beq
\label{ddeformed}
\llangle\varphi_{a_1}\cdots\varphi_{a_n}\rrangle_{\lambda^i}
=\llangle e^{-\int\lambda^i t_i\, d^p\tau}\varphi_{a_1}\cdots\varphi_{a_n}\rrangle_{0}\,,
\eeq
for a $p$-dimensional defect.
In particular for a pair of $\phi=t_i$ we have
\beq
\label{dd2pt}
\llangle t_{i}t_{j}\rrangle_{\lambda^i}
=\llangle e^{-\int\lambda^i t_i\, d^p\tau} t_{i} t_{j}\rrangle_{0}\,.
\eeq
This is the extension of the local Zamolodchikov metric \eqref{Nt}
beyond the flat space approximation 
and the derivatives with respect to $\lambda^i$ give the Riemann tensor. Indeed, 
as in \cite{Kutasov:1988xb}, one finds
\beq
\label{double-integral}
R_{ijkl}=\int d^p\tau_1 d^p\tau_2 
\Big[\llangle t_j(\tau_1) t_k(\tau_2) t_i(0) t_l(\infty)\rrangle_c
-\llangle t_j(0)t_k(\tau_2) t_i(\tau_1) t_l(\infty)\rrangle_c\Big]\,,
\eeq
where $\llangle\dots\rrangle_c$ implies the connected correlator, 
as stressed for example in \cite{Friedan:2012hi}.
This integral is $2p$ dimensional, but it can be reduced to an integral over 
cross-ratios 
\cite{Friedan:2012hi,Drukker:2022dzs}.

Given that the manifold is $G/G'$, there is no mystery in the metric. Indeed if it is a maximally 
symmetric space, the Riemann tensor is determined by the Ricci scalar $R$ as
\beq
\label{curv-symmetric}
R_{ijkl}=\frac{R}{D(D-1)}(g_{ik}g_{jl}-g_{il}g_{jk})\,.
\eeq
where $D$ is the dimension of the conformal manifold. 
If we know the exact value of $\mathcal{C}_t$, then we know the exact form of the curvature 
and equating the last two equations 
gives a non-trivial relation on the four point function \cite{Drukker:2022dzs}.
\subsection{$O(N)$ Model}
In the $O(N)$ model with a line defect the conformal manifold is
\begin{align}
    S^{N-1}\cong O(N)/O(N-1) \,. 
\end{align}
As discussed in section \ref{Analytic}, we know the normalization of the two point function of the tilts, it is given by
\begin{align}
    \mathcal{C}_t^2=\mathcal{N}_t^2=h^2\kappa\Big(1-\frac{(1+\gamma_E+\log\pi)\epsilon}{2}\Big)+O(\epsilon^2) \,.
\end{align}
We can use then equations \eqref{double-integral} and \eqref{curv-symmetric} along with the four point function of tilts in the $O(N)$ defect model \eqref{4ptSAT} and find
\begin{align}
\label{constraint}
    \int_{0}^{1} \frac{d\xi}{\xi^2} \log\xi\left(\frac{(N+1)}{N-1}G_{At}^T(\xi)-2G_{At}^S(\xi)\right)=\frac{1}{2\mathcal{N}_t^2} \,.
\end{align}
There is a subtlety in this integral that we will investigate now. Recall that the lowest lying operator in the singlet channel of tilts is the operator $\phi_1$, this operator has scaling dimension given in \eqref{phi_1_sclaing}, and the contribution of this operator to the integral is given by 
\begin{align}
    -2\lambda_{tt\phi_1}^2\int_{0}^{1} d\xi\log(\xi)\xi^{\Delta_{\phi_1}-2}=\frac{2\lambda_{tt\phi_1}^2}{(\Delta_{\phi_1}-1)^2}\,. 
\end{align}
Since $\Delta_{\phi_1}=1+\epsilon + O(\epsilon^2)$ we notice that this term will mix orders in the $\epsilon$-expansion. To avoid complications of this we extract this contribution from the integral and then remove it from both sides of equation \eqref{constraint}, the result is 
\begin{align}
\label{constraint_cft}
     \int_{0}^{1} \frac{d\xi}{\xi^2} \log\xi\left(\frac{(N+1)}{N-1}G_{At}^T(\xi)-2G_{At}^S(\xi)\right)\Big |_{\text{no mixing}}=\frac{1}{2\mathcal{N}_t^2}-\frac{2\lambda_{tt\phi_1}^2}{(\Delta_{\phi_1}-1)^2} \,.
\end{align}
Substituting now the results \eqref{result_1_g},\eqref{result_2_g} of the analysis performed for the general ansatzes in subsection \ref{general_ansatz}, we find that the integrand vanishes at order $\epsilon$, therefore we can't use this constraint to find the parameter $a_T$. This should be expected as we can observe from equation \eqref{4pt_tilt} that the four point function of tilts at order $\epsilon$ is totally symmetric and depends only on one function, namely $I(\xi)$. Therefore, the anti-symmetrization of the four point function in the curvature tensor should vanish. However, we can use this result to predict the order $\epsilon^2$ correction to the $\lambda_{tt\phi_1}$ OPE coefficient. First, we examine up to order $\epsilon^0$ the constraint, for this let us write the OPE coeffecient as
\begin{align}
    \lambda_{tt\phi_1}=\sum_{m=0}^{\infty}\lambda_{tt\phi_1}^{(m)}\epsilon^m\,.
\end{align}
The CFT data we need are given in \eqref{Couplings_fixed_point},\eqref{phi_1_sclaing},\eqref{Normalizations}, substituting in the constraint \eqref{constraint_cft} we get
\begin{align}
    0=\frac{2\pi^2}{(N+8)}-\frac{2(\lambda_{tt\phi_1}^{(0)})^2}{\epsilon^2}-\frac{4\lambda_{tt\phi_1}^{(0)}\lambda_{tt\phi_1}^{(1)}}{\epsilon}-2(\lambda_{tt\phi_1}^{(1)})^2 \,.
\end{align}
Solving the above equation gives the results
\begin{align}
    \lambda_{tt\phi_1}^{(0)}=0,\quad (\lambda_{tt\phi_1}^{(1)})^2=\frac{\pi^2}{N+8}\,.
\end{align}
This matches the results derived from Feynman diagrams in \cite{Gimenez-Grau:2022czc}.

We are now ready to predict the $\epsilon^2$-order correction to the OPE coefficient. Using again the CFT data in \eqref{Couplings_fixed_point},\eqref{phi_1_sclaing},\eqref{Normalizations} and the constraint \eqref{constraint_cft} we find
\begin{align}
    \lambda_{tt\phi_1}^{(2)}=-\frac{(494+127N+9N^2)\pi}{4(N+8)^{\frac{5}{2}}}\,.
\end{align}
To our knowledge this is a result that doesn't exist in the literature so far, which makes the above constraint a very useful tool. Thus, we consider next the chiral Heisenberg model and the $O(n)\times O(m)$ model since in these models the tilt four-point functions have the same analytic form as the $O(N)$ model, and we present the analogous constraint.
\subsection{Chiral Heisenberg Model}
The chiral Heisenberg model action is given by
\begin{align}
    S=\int d^dx \Big(\frac{1}{2}\partial_{\mu}\Phi_I\partial^{\mu}\Phi_I+i\Bar{\psi}(\mathbf{1}_{2}\otimes\gamma^{\mu})\partial_{\mu}\psi+y\Phi_I\Bar{\psi}(\sigma_I\otimes\mathbf{1}_{2N_f})\psi+\frac{1}{8}\lambda(\Phi^2)^2\Big) \,.
\end{align}
It contains three real scalars $\Phi_I$ with $I=1,2,3$, $N_f$ Dirac fermions and $\sigma_I$ are the Pauli matrices. The theory has $SO(3)$ global symmetry and it is conformal.

In \cite{Andy} the beta functions for the couplings $\lambda$ and $y$ were calculated and the fixed points found were
\begin{align}
    y^2=\frac{1}{N+2}\epsilon+O(\epsilon^2),\quad \lambda=\frac{\sqrt{S_N}-N+2}{22(N+2)}\epsilon+O(\epsilon^2)\,,
\end{align}
with $N=4N_f$ and $S_N=N^2+172N+4$. 

Adding now a defect deformation of the form
\begin{align}
    S_{cH}\xrightarrow{}S_{cH}+h_{cH}\int d\tau\Phi_1(\tau) \,,
\end{align}
we break $SO(3)$ symmetry of the theory to $SO(2)$, therefore the defect conformal manifold is given by 
\begin{align}
    S^2\cong SO(3)/SO(2)\,.
\end{align}
Due to the breaking of $SO(3)$ we have two exactly marginal operators, the tilts $t_{\hat{I}}$, $\hat{I}=1,2$, they have protected scaling dimension $\Delta_t=1$. The operator $\Phi_1$ that couples to the defect has scaling dimension given by \cite{Andy}
\begin{align}
    \label{scaling_chiral_phi_1}
    \Delta_{\Phi_1}=1+\epsilon\frac{2}{N+2}+O(\epsilon^2)\,.
\end{align}
The beta function for the defect coupling $h_{cH}$ was also calculated in \cite{Andy} and the fixed point for the dCFT is given by
\begin{align}
    \label{fixed_point_chiral}
    h_{cH}^2=\frac{44}{\sqrt{S_N}-N+2}+O(\epsilon)\,.
\end{align}
In this model the four point function of the tilts will again be totally symmetric and depend only on the same function $I(\xi)$, this can be understood from the action of the theory and the interactions in it. Hence, we can deduce that an analogous constraint as \eqref{constraint_cft} should exist. This constraint reads in this case
\begin{align}
    \int_{0}^{1} \frac{d\xi}{\xi^2}\log\xi \Big(\frac{N+1}{N-1}G_{cHt}^T(\xi)-2G_{cHt}^S(\xi)\Big)\Big|_{\text{no mixing}}=\frac{1}{2\mathcal{C}_{cHt}^2}-\frac{2\lambda_{tt\Phi_1}^2}{(\Delta_{\Phi_1}-1)^2}\,.
\end{align}
The knowledge of the CFT data in this theory and of the normalization of tilts $\mathcal{C}_{cHt}$ doesn't allow us to predict an $\epsilon^2$-order correction to the OPE coefficient $\lambda_{tt\Phi_1}$, but still we can use the constraint to derive the order $\epsilon^0$ and $\epsilon$ correction. This works as follows, first the normalization is given by $\mathcal{C}_{cHt}^2=h_{cH}^2\kappa$, and the scaling dimension of $\Phi_1$ is given in \eqref{scaling_chiral_phi_1}, we use these and find
\begin{align}
    0=\frac{\pi^2(\sqrt{S_N}-N+2)}{22}-(N+2)^2\frac{(\lambda_{tt\Phi_1}^{(0)})^2}{2\epsilon^2}-(N+2)^2\frac{\lambda_{tt\Phi_1}^{(0)}\lambda_{tt\Phi_1}^{(1)}}{\epsilon}-(N+2)^2\frac{(\lambda_{tt\Phi_1}^{(1)})^2}{2}+O(\epsilon) \,.
\end{align}
This gives
\begin{align}
    \lambda_{tt\Phi_1}^{(0)}=0,\quad  (\lambda_{tt\Phi_1}^{(1)})^2=\frac{\pi^2(\sqrt{S_N}-N+2)}{11(N+2)^2}\,.
\end{align} 
\subsection{$O(n)\times O(m)$ Model}
\label{sec:integral}
In this section we discuss the analogous constraint \eqref{constraint_cft} in the case of the $O(n)\times O(m)$ model. The action for this model reads
\begin{align}
    S=\int d^dx \Big( \frac{1}{2}\partial_{\mu}\hat{\phi}_{\hat{i}}\partial^{\mu}\hat{\phi}_{\hat{i}}+\frac{1}{2}\partial_{\mu}\check{\phi}_{\check{i}}\partial^{\mu}\check{\phi}_{\check{i}}+\frac{1}{8}\lambda_1(\hat{\phi}^2)^2+\frac{1}{8}\lambda_2(\check{\phi}^2)^2+\frac{1}{4}g(\hat{\phi}^2\check{\phi}^2\Big) \,.
\end{align}
The $\hat{\phi}_{\hat{i}}$ fields transform in the fundamental of $O(n)$ with $\hat{i}=1,\dots,n$, and the $\check{\phi}_{\check{i}}$ fields transform in the fundamental of $O(m)$ with $\check{i}=1,\dots,m$. We are going to consider the case where $m=n$, the fixed point then is given by \cite{Andy}
\begin{align}
\label{couplings_fixed_O(n)xO(m)}
    \lambda_1=\lambda_2=\frac{n}{2(n^2+8)}\epsilon,\quad g=-\frac{n-4}{2(n^2+8)}\epsilon \,,
\end{align}
where this is a stable fixed point for $2<n<4$.

Let us consider now the following defect deformation
\begin{align}
    S\xrightarrow{}S+\hat{h}\int d\tau \hat{\phi}_{\hat{1}}+\check{h}\int d\tau \check{\phi}_{\check{1}} \,,
\end{align}
with $\hat{\phi}_{\hat{1}}$, $\check{\phi}_{\check{1}}$ having scaling dimensions
\begin{align}
\label{scaling_dimensions_O(n)xO(m)0}
\Delta_{\hat{\phi}_{\hat{1}}}=\Delta_{\check{\phi}_{\check{1}}}=1+\epsilon+O(\epsilon^2)\,.
\end{align}
the authors of \cite{Andy} calculated the beta functions for the defect couplings $\hat{h},\check{h}$ and found the defect CFT fixed points at 
\begin{align}
\label{h_fixed_point_O(n)xO(m)}
    \hat{h}^2=\frac{g-\lambda_1}{g^2-\lambda_1\lambda_2}\epsilon,\quad \check{h}^2=\frac{g-\lambda_2}{g^2-\lambda_1\lambda_2}\epsilon \,.
\end{align}
The symmetry breaking pattern is that of $O(n)\times O(m)\xrightarrow{} O(n-1)\times O(m-1)$, and hence the conformal manifold is given by
\begin{align}
    \frac{O(n)\times O(m)}{O(n-1)\times O(m-1)}=S^{n-1}\times S^{m-1}\,.
\end{align}

This means that there are two classes of tilt operators, the first one transforms in the fundamental of $O(n-1)$ and reads $\hat{t}_{\hat{I}}$, with $\hat{I}-1,\dots,n-1$, and the second one transforms in the fundamental of $O(m-1)$ and reads $\check{t}_{\check{I}}$, with $\check{I}=1,\dots,m-1$.

To proceed, we can restrict ourselves to the study of the relevant constraint for $\hat{t}_{\hat{I}}$, and the other one follows in exactly the same way. One notable difference in this case, is the interaction term $g(\hat{\phi}^2\check{\phi}^2)$ that allows for mixed correlators, a consequence of this, is that the singlet OPE of the tilts in the $O(n-1)$ group will be
\begin{align}
\label{OPE_O(n)_O(m)}
(\hat{t}_{\hat{I}}\times\hat{t}_{\hat{J}})_S=\hat{\phi}_{\hat{1}}+\check{\phi}_{\check{1}}+\dots\,.
\end{align}
We notice that both the defect operators that couple to the line appear in the singlet OPE $(\hat{t}_{\hat{I}}\times\hat{t}_{\hat{J}})_S$.

As far as the four point function is concerned, once again, we can conclude that based on the quartic interactions in the action, the four point function of the tilts in $O(n-1)$ will be totally symmetric at order $\epsilon$ and thus the integral will vanish. However, as mentioned, we need to be careful when removing the operators $\hat{\phi}_{\hat{1}}$ and $\check{\phi}_{\check{1}}$ from the integral, as now we need to remove both, because they appear in the singlet OPE as we discussed. This results into the following integral identity
\begin{align}
\label{constraint_O(n)xO(m)}
    \int_{0}^{1}\frac{d\xi}{\xi^2}\log\xi\Big(\frac{n+1}{n-1}G_{\hat{t}}^T(\xi)-2G_{\hat{t}}^S(\xi)\Big)\Big|_{\text{no mixing}}=\frac{1}{2\mathcal{C}_{\hat{t}}^2}-\frac{2\big(\lambda_{\hat{t}\hat{t}\hat{\phi}_{\hat{1}}}+\lambda_{\hat{t}\hat{t}\check{\phi}_{\check{1}}}\big)^2}{\big(\Delta_{\hat{\phi}_{\hat{1}}}-1\big)^2}\,.
\end{align}
This can serve now as a constraint on the CFT data. After using the constraint along with equations \eqref{scaling_dimensions_O(n)xO(m)0},\eqref{couplings_fixed_O(n)xO(m)},\eqref{h_fixed_point_O(n)xO(m)} and the fact that the integrand is zero at order $\epsilon$, we find the following results 
\begin{align}
    \lambda_{\hat{t}\hat{t}\hat{\phi}_{\hat{1}}}^{(0)}= \lambda_{\hat{t}\hat{t}\check{\phi}_{\check{1}}}^{(0)}=0, \quad \lambda_{\hat{t}\hat{t}\hat{\phi}_{\hat{1}}}^{(1)}+\lambda_{\hat{t}\hat{t}\check{\phi}_{\check{1}}}^{(1)}=\pm\frac{\pi}{\sqrt{2(n^2+8)}}\,.
\end{align}
We notice in this case that we can only fix a linear combination of the order $\epsilon$ correction of the two OPE coefficients.

Finally, to all three cases we considered here, in order to predict higher corrections to the OPE coefficient, we need higher corrections to the rest of the CFT data as well. Also, After order $\epsilon^3$ we need also the integral part, for this we would need the $\epsilon^2$ correction of the four-point function as well.
\section{Conclusions and Outlook}
In this work we've introduced an application of the identities in  \cite{Kalmykov:2006pu} that is used to calculate CFT data in defect conformal field theories. We took advantage of relations between hypergeometrics presented in \cite{Kalmykov:2006pu}, and arrived at a recursion relation that allows one to compute the hypergeometric $_2F_1(n+\gamma\varepsilon,n+\gamma\varepsilon;2n+2\gamma\varepsilon;\xi)$ for any $n$, exactly in the cross ratio $\xi$, and at order $\epsilon$. We can use this result to expand the resulting hypergeometric in series, and match order by order with the explicit four point function result, inverting in this way the four point function, and deriving the CFT data. The advanatage of this algorithmic procedure is that it is very simple to implement with code, and that it requires less time to work in contrast with Mathematica packages that expand hypergeometrics \cite{package1,package2}. As a result of this speed up in the calculation time, one can deduce CFT data for operators up to $n=10$ in a very short amount of time. This was almost impossible to do with the Mathematica packages. As an example we've inverted four point functions of tilts and displacements in the $O(N)$ defect model. In these calculations we've used explicit results for the four point function that were found in \cite{Gimenez-Grau:2022czc}. We were able to derive averaged CFT data for the singlet, traceless-symmetric and anti-symmetric channels for up to $n=10$ for the first two channels and $n=11$ for the last. The reason we computed averaged CFT data is operator mixing, we shed some light on this in Appendix \eqref{Appendix_A}. We were also able to resolve mixing for the simple case of $n=2$ in the singlet channel of tilts in the $O(N)$ model. Furthermore, we performed a general ansatz analysis for the four point function of tilts in the $O(N)$ model with a line defect, and found that we can restrict the ansatz up to one undetermined parameter. Lastly, we've explored the consequences of this to the integral identities of \cite{Drukker:2022dzs}. We've found that we cannot determine the undetermined parameter, but we arrived at a constraint that includes defect CFT data, and we can use it to predict corrections to OPE coefficients of the theory. Specifically, in the $O(N)$ model we could predict the $\epsilon^2$ correction to the $\lambda_{tt\phi_1}$ OPE coefficient. Then, we've explored the analogous identity for the chiral Heisenberg model and for the $O(n)\times O(m)$ model. The results we arrived at were order $\epsilon^0$ and $\epsilon$ corrections to the OPE coefficients. Looking further ahead, it would be an interesting direction to explore if a similar analysis can be performed at order $\epsilon^2$, this would require results for the four point functions of tilts and displacements at order $\epsilon^2$. Or, one could also explore general ansatz solutions to order $\epsilon^2$ and restrict it similarly to the order $\epsilon$ calculation presented here. Finally, improving the algorithm for inverting or calculating with the recursion the hypergeometrics, could lead to an improved time of obtaining the CFT data and raise the $n$ much higher than $n=10$. We wish to return on these matters in future work. 
\section*{Acknowledgements}
I would like to thank Nadav Drukker, Andreas Stergiou and Ziwen Kong for the very useful, insightful and enlightening discussions. My research is funded by STFC Grant No. ST/W507556/1..
\appendix
\section{Averaged CFT data}
\label{Appendix_A}
In this appendix we discuss the relation between the actual CFT data of the operators that mix at order $\epsilon$ and the averaged CFT data that we are calculating. We will explicitly do so for the case of the operators $S1,S2$ with classical scaling dimension $\Delta_{S1}=\Delta_{S2}=2$ in the $O(N)$ model with a line defect. The result is 
\begin{align}
    &(\lambda_{ttS}^{(0)}[2])^2=(\lambda_{ttS1}^{(0)})^2+(\lambda_{ttS2}^{(0)})^2\,,\\
    &(\lambda_{ttS}^{(1)}[2])^2=\frac{\lambda_{ttS1}^{(0)}\lambda_{ttS1}^{(1)}+\lambda_{ttS2}^{(0)}\lambda_{ttS2}^{(1)}}{\sqrt{(\lambda_{ttS1}^{(0)})^2+(\lambda_{ttS2}^{(0)})^2}}\,,\\
    &\gamma_{S}[2]=\frac{\gamma_{S1}(\lambda_{ttS1}^{(0)})^2 + \gamma_{S2} (\lambda_{ttS2}^{(0)})^2}{(\lambda_{ttS1}^{(0)})^2+(\lambda_{ttS2}^{(0)})^2} \,.
\end{align}
Similar expressions can be found for higher dimension operators, one would have to know the exact number of mixed operators per scaling dimension in the theory to perform this calculation, we leave this for future work.
\section{$1d$ Conformal Blocks}
\label{conformal_block}
In this appendix we are going to discuss the conformal blocks of a $1d$ dCFT and their relation to the hypergeometric $ _2F_1(\Delta,\Delta;2\Delta;\xi)$ (for more details on conformal blocks see \cite{Dolan&Osborn_3}).

First, let us start with a four point function of a generic defect operator $O$ in the $1d$ dCFT
\begin{align}
    \llangle O(\tau_1)O(\tau_2)O(\tau_3)O(\tau_4)\rrangle = \frac{\mathcal{N}_O^2}{\tau_{12}^{2\Delta}\tau_{34}^{2\Delta}}\sum_{\hat{\Delta}}\lambda_{OO\hat{O}}^2\hat{G}_{\hat{O}}^{(1)}(\xi) \,.
\end{align}
We have used the OPE 
\begin{align}
    O(\tau_1) \times O(\tau_2) = \sum_{\hat{\Delta}}\lambda_{OO\hat{O}}(\tau_{12},\partial_{\tau})\hat{O}_{\hat{\Delta}}(\tau)\Big|_{\tau=\tau_1}\,,
\end{align}
where we sum over all the exchanged $\hat{O}_{\hat{\Delta}}$ defect primary operators. 

In order to find the conformal block $\hat{G}_{\hat{O}}^{(1)}(\xi)$ we will use the fact that primaries $\hat{O}_{\hat{\Delta}}$ are eigenstates of the Casimir of the $1d$ conformal group $SO(2,1)$, satysfying
\begin{align}
    \hat{C}^2 \hat{O}_{\hat{\Delta}}=\hat{\Delta}(\hat{\Delta}-1)\hat{O}_{\hat{\Delta}}\,.
\end{align}
Combining the above with conformal invariance and inserting the representation of the identity $\sum_{\hat{\Delta}}|\hat{O}_{\hat{\Delta}}\rangle\langle\hat{O}_{\hat{\Delta}}|$, we arrive at the following differential equation for the $1d$ conformal block $\hat{G}_{\hat{O}}^{(1)}(\xi)$
\begin{align}
    \mathcal{D}\hat{G}_{\hat{O}}^{(1)}(\xi)=\hat{\Delta}(\hat{\Delta}-1)\hat{O}_{\hat{\Delta}}\,,
\end{align}
with the differential operator $\mathcal{D}$ given by 
\begin{align}
    \mathcal{D}=\xi\Big(-\xi\frac{d}{d\xi}-(\xi-1)\xi\frac{d^2}{d\xi^2}\Big)\,.
\end{align}
The solution to the above differential equation is the well known $1d$ conformal block given by
\begin{align}
    \hat{G}_{\hat{O}}^{(1)}(\xi)=\xi^{\hat{\Delta}} {\,}_2F_1(\hat{\Delta},\hat{\Delta};2\hat{\Delta};\xi) \,.
\end{align}
\bibliographystyle{utphys2}
\bibliography{refs}
\end{document}